# Long-lived direct and indirect interlayer excitons in van der Waals heterostructures

Bastian Miller,[1,2] Alexander Steinhoff,[3] Borja Pano,[1,2] Frank Jahnke,[3] Alexander Holleitner,[1,2]* Ursula Wurstbauer[1,2]*

[1] Walter Schottky Institut and Physics-Department, Technical University of Munich, Am Coulombwall 4a, 85748 Garching, Germany.

[2] Nanosystems Initiative Munich (NIM), Schellingstr. 4, 80799 München, Germany.

[3] Institut für Theoretische Physik, Universität Bremen, P.O. Box 330 440, 28334 Bremen, Germany

**Abstract**

We investigate the photoluminescence of interlayer excitons in heterostructures consisting of monolayer $MoSe_2$ and $WSe_2$ at low temperatures. Surprisingly, we find a doublet structure for such interlayer excitons. Both peaks exhibit long photoluminescence lifetimes of several ten nanoseconds up to 100 ns at low temperatures, which verifies the interlayer nature of both. The peak energy and linewidth of both show unusual temperature and power dependences. In particular, we observe a blue-shift of their emission energy for increasing excitation powers. At a low excitation power and low temperatures, the energetically higher peak shows several spikes. We explain the findings by two sorts of interlayer excitons; one that is indirect in real space but direct in reciprocal space, and the other one being indirect in both spaces. Our results provide fundamental insights into long-lived interlayer states in van der Waals heterostructures with possible bosonic many-body interactions.

Semiconductor heterostructures (HS) are very often the foundation for the observation of novel phenomena in both fundamental science as well as device applications. The electrical and optical properties of such HS can be engineered in a wide range resulting in precisely tailored functionalities. Of particular interest are optically active HS facilitating many device applications such as photo-detectors, solar cells, light-emitting diodes or lasers and fostering the observation of many-body driven quantum phenomena found in systems with reduced dimensionality such as quantum wells or quantum dots. Excitons are electron-hole pairs coupled by attractive Coulomb interaction which results in a ground state with a reduced energy compared to the corresponding single-particle energies. Exciton ensembles exhibit an intriguing interaction driven phase diagram with different classical and quantum phases including quantum liquids and solids [1–3]. The bosonic nature of excitons enables these composite particles even to condensate into macroscopic ground state wave functions forming a Bose-Einstein condensate (BEC) [4].

Ensembles of indirect a.k.a. interlayer excitons (IXs) are particularly fascinating systems to explore such classical and quantum phases of interacting bosonic ensembles. IXs are composite bosons that feature enlarged lifetimes due to the reduced overlap of the electron-hole wave functions resulting in dense IX ensembles that are thermalized to the lattice temperature. Besides IX ensembles in III-V HS [5–15], hetero-bilayers prepared from semiconducting transition metal dichalcogenides (TMDs) exhibit superior potential for studying interacting IX ensembles [3,16–25] with intriguing spin- and valley-properties [26–28]. At the same time, TMDs are truly two-dimensional (2D) crystals coupled to each other or to substrates by van der Waals forces. Furthermore, 2D materials provide a unique platform to stack them into HS without the limitations of lattice mismatch in conventional crystals as e.g. GaAs and InAs. Semiconducting TMDs such as $MoSe_2$ and $WSe_2$ that have a direct band gap in the monolayer limit [29–31] excel on further properties due to a strong light-matter interaction [20] with an absorbance of up to 15% of visible light in just one monolayer [32–34]. They possess also fascinating spin- and valley-properties with both degrees of freedom locked by the lack of an inversion center in the monolayer crystals enabling optical valley polarization [35–38].

The electronic band structure of TMDs is not only altered by the number of layers [29–31] but also by doping [39–42]. Increasing the charge carrier density either by doping or intensive irradiation with light above the bandgap results in a reduction of the electronic bands at the K- and K'-points at the zone boundary [39–43]. An even larger renormalization effect of the conduction band occurs for the $\Sigma$-valley in the middle of the Brillouin zone between the $\Gamma$- and K-points such that the monolayer system can turn into an indirect semiconductor [40,41]. The doping induced lowering in the band gap and the reduced exciton binding energies nearly cancel out each other so that only the energy of the neutral exciton in photoluminescence (PL) experiments remains - almost unaffected by doping. However, the PL intensity is reduced by doping until the exciton Mott transition is achieved e.g. by high illumination intensities [42,43].

A further intriguing property of semiconducting TMDs is a strong exciton binding of up to 0.5 eV [44,45] that is one order of magnitude more than in conventional III-V materials systems. This can be traced back to the fact that the dielectric screening is strongly reduced in these less than 1 nm thin crystals. For this reason, excitonic phenomena are expected at rather high temperatures up to room temperature in 2D monolayers. Moreover, the exciton Bohr radius of only a few lattice sites is expected to enable the creation of rather dense exciton ensembles - a rigorous requirement for studying the quantum phase diagram of interacting exciton ensembles [1]. Moreover, van der Waals heterostructures feature a type-II band alignment [16,19,25,46], and therefore, they render an efficient transfer of photogenerated charge carriers such that electrons accumulate in one layer and holes in the other layer [23,27,47]. In turn, thermalized IXs are expected with a significantly enhanced IX lifetime and density [21,22,26]. Furthermore, by optical spin-valley pumping in one layer, the spin- and valley- degrees of freedom are preserved by the interlayer transfer of polarized charge carriers [25,27,28,48]. While intralayer excitons in TMD monolayers are perfect in-plane dipoles fostering Förster-type energy transfer between hetero-bilayers [49], interlayer excitons possess a permanent out of plain component altering the coupling to light and enabling to tune the exciton properties by static electric fields perpendicular to the layers. The exciton binding energies of IXs is reduced in HS from about 0.5 eV for the individual layers to about 0.2 eV in the HS [25], but it is still one order of magnitude larger compared to IX in III-V HS.

In TMD-based heterostructures, the mismatch in the lattice constants of the two constituting monolayers gives rise to a Moiré pattern, which can impact the electronic band structure [50]. For $MoS_2$/$WSe_2$ heterostructures, a lateral modulation of the interlayer bandgap of up to ~15 meV was reported [50]. In particular, the difference of $MoS_2$-CB minima at the $\Sigma$- and K -points is sizeable modulated. In turn, the lattice mismatch between the two layers of the heterostructure can cause a displacement of the conduction band minimum and valence band maximum at K and K', but it also affects other high symmetry points [48]. This is of relevance for optoelectronic exciton circuits and also for their lateral confinement.

Here, we investigate the properties of dense and diluted interlayer exciton (IX) ensembles in a van der Waals HS prepared from monolayers of $MoSe_2$, and $WSe_2$ by a comprehensive photoluminescence study. We characterize the power-, temperature-, and polarization- dependent photoluminescence both in time-integrated and time-resolved experiments. Most importantly, we detect an extra red-shifted luminescence signal with a significantly increased lifetime. Both characteristics are a strong indication for the formation of IXs in the HS. Surprisingly, we find that this IX emission is composed of two peaks with different lifetimes as well as with distinct power- and temperature-dependences. We assign one transition to be indirect in real space, however, direct in momentum space and the second one to be indirect in both momentum and real space. The observed homogenous filling of large lateral areas of the van der Waals HS together with the long IX lifetimes and interesting polarization behavior makes van der Waals HS highly attractive to study correlated quantum phases such as bosonic liquids or a Bose-Einstein condensation in these composite bosonic particle ensembles, but also for the design of novel optoelectronic device architectures.

The HS is a hetero-bilayer, which is prepared by micromechanical exfoliation from bulk crystals. The top monolayer of $WSe_2$ is stacked onto the lower $MoSe_2$ using a viscoelastic dry stamping method [51,52], such that the crystal axes are rotationally aligned with a precision of about $\pm1°$. The HS are heated in vacuum (<$10^{-2}$mbar, 150°C, 12h) to remove trapped residues in the van der Waals gap between the two crystals and to enlarge the coupling between the layers [53]. The HS is placed on optically inactive, ultra-smooth glass substrates made from Schott borofloat®33. An optical microscope image of the described $MoSe_2$/$WSe_2$ hetero-bilayer is plotted in Figure 1(a). The individual monolayers as well as the region of the HS are marked. We use Raman spectroscopy, atomic force microscopy and µ- photoluminescence (PL) spectroscopy to characterize the samples. Figure 1(b) displays a typical low temperature ($T_{bath}$ = 3K) PL spectrum showing the emission of the intralayer excitons (A-excitons) from MoSe$_2$ at 1.65 eV and from WSe$_2$ at 1.74 eV. Strikingly, a well-pronounced, significantly red-shifted luminescence signal appears in the range of 1.33 to 1.38 eV. We identify the latter to stem from IXs, in agreement with recent reports in literature [21,22,25]. The HS is excited by a laser at an energy $E_{laser}$ = 1.94 eV and a power in the range 2 nW ≤ $P$ ≤ 140 µW with a focused spot diameter of 1 µm on the sample. The emitted luminescence is dispersed by a single grating spectrometer and recorded with a charge coupled device (CCD) camera with a combined spectral resolution of about 0.2 meV. Notably, the signal that we interpret to stem from interlayer excitons [IX in Figure 1(b)] splits into two peaks. A careful line-shape analysis reveals that the IX peak is composed of two Gaussian shaped contributions. The one at an energy of $E^{\blacktriangledown}_{IX}$ = 1.38 eV is marked by a filled triangular symbol (▼) and is referred to as IX$^{\blacktriangledown}$ in the following. The one at an energy of $E^{\triangledown}_{IX}$ 1.33 eV is defined as IX$^{\triangledown}$. We observe that the PL intensity of both intralayer excitons in MoSe$_2$ and WSe$_2$ is reduced compared to the integral intensity of IX$^{\blacktriangledown}$ and IX$^{\triangledown}$. This bleaching of the direct excitons indicates an efficient charge transfer across the heterojunction [47]. In particular, the dissociation of the hot electron-hole pairs at the junction can be assumed to be faster than the formation of intralayer excitons. In this picture, a substantial excess number of electrons accumulate in MoSe$_2$ as well as holes in WSe$_2$ within the area of the HS. These excess charge carriers can form IXs due to the attractive Coulomb interaction. When formed, a vertical separation of the bound electrons and holes gives rise to a finite dipole moment of the IXs. Additionally, the separation reduces the overlap of the IX wave function, which results in an increased lifetime of the IXs compared to the lifetimes of intralayer excitons. We would like to mention that under the excitation with circular polarized light, the high-energy peak IX$^{\blacktriangledown}$ is weakly co-polarized of about 20% (for $T_{bath}$ = 3 K , $P$ = 140 µW), whereas the low-energy peak IX$^{\triangledown}$ is unpolarized within the given noise level. Generally, the valley- and spin-polarization can be assumed to be preserved for charge carriers transferred across heterojunctions [27]. Therefore, we assign the observed polarization to the fact that our excitation energy of $E_{laser}$ = 1.94 eV is in resonance with the (intralayer) B exciton transition in MoSe$_2$ at higher temperatures and slightly below that value at low temperatures resulting in the generation of valley-polarized charge carriers in MoSe$_2$.

To further characterize the HS, we compute the essential single-particle band diagrams. The left panel in Figure 1(c) shows the conduction band (CB) of MoSe$_2$, while the right panel depicts the valence band (VB) of WSe$_2$ with and without a moderate doping density. The middle panel sketches the corresponding type-II band alignment of MoSe$_2$ and WSe$_2$ [16,19,25,46]. To calculate band-structure renormalizations due to excited carriers, we use a self-energy in screened-exchange-Coulomb-hole approximation based on material-realistic G$_0$W$_0$-band structures and Coulomb matrix elements as described in [39]. Moreover, the spin-orbit induced spin-splitting is taken into account along the lines of [39,54] including first- and second-order effects For both MoSe$_2$ and WSe$_2$, direct optical transitions can occur at the K- and K'-points [cf. closed triangle in the left panel of Figure 1(c)]. However, there is another minimum in the MoSe$_2$-CB, which is typically referred to as the Σ-valley. It is located between the Γ- and the K-point [cf. open triangle in the left panel of Figure 1(c)]. Generally, carrier doping has an impact on all bands. In particular, the minima at the Σ-valley and the K- and K'-points are lowered with an increasing charge carrier density. In our calculations, the energy of the Σ-valley is presumably below the one of the K- and K'-points. This situation turns MoSe$_2$ into an indirect band-gap material in agreement with recent reports in literature and similar findings for $MoS_2$ [40–42].

Along this first discussion, we interpret the observed doublet structure in the IX emission as two different kinds of interlayer excitons. The low-energy peak $IX^\nabla$ is interpreted to be indirect in momentum space with the electron in the $\Sigma$-valley and the hole at K- or K'-point (i.e. a $\Sigma^{MoSe2}$-$K^{WSe2}$ transition). Accordingly, we call this exciton the indirect interlayer exciton $IX^\nabla$. We interpret the high-energy exciton $IX^\blacktriangledown$ to be rather direct in momentum space with electrons and holes at the K- and K'-points, i.e. a $K^{MoSe2}$-$K^{WSe2}$ transition. Accordingly, we call it the direct interlayer exciton $IX^\blacktriangledown$. In principle, two further interpretations might explain a splitting of interlayer excitons. One would be the formation of an interlayer trion, which is a tri-particle complex with an interlayer exciton bound to either an electron in the $MoSe_2$-CB or a hole in the $WSe_2$-VB. The other potential interpretation would be the binding of electrons with opposite spin due to the spin-splitting of the $MoSe_2$-CB induced by strong spin-orbit coupling (SOC) [55]. In the following, we show a comprehensive set of PL data of the $IX^\nabla$ and $IX^\blacktriangledown$. From the corresponding discussion, we draw the conclusion that the most likely interpretation for the splitting of the IXs is due to the formation of the just-defined indirect and direct interlayer excitons.

Figure 2 shows PL measurements of the $IX^\nabla$ and $IX^\blacktriangledown$ peaks for excitation intensities varied over 5 orders of magnitude and for a bath temperature in the range of $T_{bath}$ = 3K and 70K. Figures 2(a) and 2(b) depict the original spectra in a logarithmical waterfall representation. Fitting the two emission peaks $IX^\nabla$ and $IX^\blacktriangledown$ by two Gaussians for the whole set of parameters, we extract the peak intensities $I(IX^\nabla)$ and $I(IX^\blacktriangledown)$, the emission energies $E(IX^\nabla)$ and $E(IX^\blacktriangledown)$, and the corresponding full widths at half maximum (FWHM) [cf. Figures 2(c) - 2(h)].

Intriguingly, the low-energy peak $IX^\nabla$ dominates the overall luminescence at a low laser power, and its intensity $I(IX^\nabla)$ outnumbers $I(IX^\blacktriangledown)$ by up to a factor of ten in this regime. In contrast, the luminescence of the high-energy peak $IX^\blacktriangledown$ dominates the high laser power regime. The cross-over occurs at a typical laser power of a few µW with a clear tendency of an increasing cross-over power for an increasing bath temperature [cf. dashed lines in Figure 2(c) with 2(f)]. As can be seen in Figures 2(d) and 2(g), the emission energies of the $IX^\nabla$ and $IX^\blacktriangledown$ show a clear blue-shift for higher laser powers at all temperatures. It is worth mentioning that this finding agrees well with a recent report by Nagler *et al.* [21]. However, the authors only report on the elevated power range ($P > 1$ µW) and they consider only one emission peak. In our experiments, the emission from the high-energy $IX^\blacktriangledown$ also dominates the overall emission for the highest powers. However, the $IX^\nabla$ peak is still observable also in this regime, and it even dominates the low power regime. We observe an energy difference for the peaks of about 30-50 meV [Figures 2(d) and 2(g)].

In our experiment, the linewidth of the luminescence is generally larger for the $IX^\nabla$ peak compared to the one of the $IX^\blacktriangledown$ peak [Figures 2(e) and 2(h)]. In first order, the FWHM of the $IX^\nabla$ peak has a value of ~50 meV for most of the investigated laser powers at 3K, while it increases monotonously up to a value of ~60 meV at 30K for the highest laser powers. In contrast, the FWHM of the $IX^\blacktriangledown$ seems to increase for both temperatures as a function of laser powers with maximum values of ~45 meV for the investigated range of laser powers. In addition, several rather narrow emission peaks with a FWHM of about 3 meV appear in the spectral range of the $IX^\blacktriangledown$ exclusively for an excitation power below $P \leq$ 200 nW and low temperatures $T < 30$ K [arrows in Figure 2(a)]. We ascribe them to local minima in the potential landscape, as will be discussed below.

Figure 3 gives insights into the temperature dependence of the IX emission for an intermediate laser power of $P = 200$ nW. Figure 3(a) shows the original data in logarithmic waterfall plots, and figures 3(b)-(d) summarize the corresponding temperature dependence of the intensity, the emission energy, and the FWHM of the peaks IX$^\nabla$ and IX$^\blacktriangledown$. Again, all extracted values are deduced from Gaussian fits to the data. Overall, the intensity of both emission peaks IX$^\nabla$ and IX$^\blacktriangledown$ decreases with increasing temperatures for all parameters. However, for this intermediate power of $P = 200$ nW, the IX$^\blacktriangledown$ peak loses intensity at a lower temperature than the IX$^\nabla$ peak. Only for the highest laser powers, the IX$^\blacktriangledown$ peak dominates for all temperatures. The emission energy of the IX$^\nabla$ peak decreases for $T_{bath} > 20$ K, while the one of the the IX$^\blacktriangledown$ peak increases. Moreover, there seems to be no simple trend in the FWHM vs temperature for both peaks at this power [cf. Figure 3(d)]. In our understanding, this reflects the occurrence of different regimes of the PL emission vs temperature, as is obvious in the original data of Figure 3(a) and as it will be discussed below.

We now turn to time- and polarization-resolved PL measurements for both IX$^\nabla$ and IX$^\blacktriangledown$ peaks. The spectra are excited with a $\sigma^+$-polarized laser pulse and a pulse duration of 85 ps. The emitted light is then passed through analyzer optics to detect either $\sigma^+$-polarized or $\sigma^-$-polarized light. The polarization resolved PL light is recorded by a time correlated avalanche photo diode (APD). The spectral resolution is achieved by two sets of filters adjusted, such that either PL from the IX$^\blacktriangledown$ or from the IX$^\nabla$ emission is recorded by the APD. Figure 4 shows the time-resolved PL signal for a circular-co-($\sigma^{++}$) and circular-cross-($\sigma^{+-}$) polarization configuration for low and high excitation power ($P = 100$ nW and $P = 10$ µW). In particular, Figures 4(a) and (b) display the polarization-resolved temporal evolution of the IX PL signal for low excitation, while the panels (c) and (d) show the identical measurements for high power excitation. The lifetime of the IX ensembles extracted from bi-exponential fits to the data constitutes several ten ns up to ~100 ns [cf. Figures 4(e) and 4(f)]. For all investigated experimental parameters, the found values are significantly larger compared to the lifetime of intralayer excitons. This observation verifies the interlayer nature of the IX$^\nabla$ and IX$^\blacktriangledown$ peaks. The values for the lifetimes coincide with values reported in literature spreading from less than 2 ns [22] to more than 180 ns [21].

Figure 5 shows the spatial distribution of the PL measured at the emission energy of the IX$^\nabla$ and IX$^\blacktriangledown$ peaks, respectively. The area coincides with the one of the HS, which is highlighted by the dashed square in Figure 1(a). Figures 5(a) and 5(b) depict corresponding intensity maps for a low excitation power ($P = 200$ nW), while Figures (c) and (d) present equivalent maps for a rather high power ($P = 100$ µW). For the latter, the PL is spatially homogeneous for both IX$^\nabla$ and IX$^\blacktriangledown$ peaks [cf. (c) and (d)], which is consistent with a homogenously filled dense ensemble of long-living interlayer excitons spread over the whole area of the HS. For the low excitation power ($P = 200$ nW), the emission from the energetically higher IX$^\blacktriangledown$ shows a more inhomogeneous spatial distribution of the PL intensity [Figure 5(b)]. This is the regime, in which we also observe the spectral spikes in the PL spectra [cf. arrows in Figure 2(a)]. In contrast, the emission of the energetically lower IX$^\nabla$ is spatially homogeneous also for this low power regime [Figure 5(a)] indicating that the potential landscape of the HS is rather smooth or smoothed out on the level of the low-energy IX$^\nabla$, whereas there are lateral fluctuations in the potential landscape for the emission from the high-energy IX$^\blacktriangledown$ peak. We note that for temperatures higher than $T_{bath} > 30$ K, the PL maps are spatially homogeneous for low and high excitation powers for both IX$^\triangledown$ and IX$^\blacktriangledown$ peaks.

To start with, the different behavior in the lateral homogeneity of the low-energy IX$^\triangledown$ and the high-energy IX$^\blacktriangledown$ peak at a low laser power and low temperatures indicates that the two emission peaks stem from two interlayer excitons of different nature. Moreover, the IX$^\triangledown$ peak shows longer lifetimes and a distinctively broader linewidth than the IX$^\blacktriangledown$ peak throughout the examined range of parameters. Moreover, the IX$^\triangledown$ dominates the luminescence at low power and the IX$^\blacktriangledown$ does so at high power.

In general, neglecting attractive Coulomb interaction between electrons and holes in adjacent layers, the single-particle band diagrams of MoSe$_2$ and WSe$_2$ [Figure 1(d)] already suggest two different kinds of interlayer excitons that are either indirect or direct in momentum space with the holes located at the K$^{WSe2}$-point of the WSe$_2$-VB and the electrons either at the $\Sigma^{MoSe2}$-valley or the K$^{MoSe2}$-points of the MoSe$_2$-CB. In such a scenario with presumably adjusted lattices of MoSe$_2$ and WSe$_2$, the lowest energy transitions of the HS would be (1) a momentum indirect interlayer transition between $\Sigma^{MoSe2}$ - K$^{WSe2}$ and (2) a momentum direct interlayer transition between K$^{MoSe2}$ - K$^{WSe2}$. Moreover, the single-particle band diagrams of Figure 1(c) predict that the momentum indirect transition $\Sigma^{MoSe2}$ - K$^{WSe2}$ is lower in energy than the one at K$^{MoSe2}$ - K$^{WSe2}$, in agreement with our experimental findings. However, the calculations do not consider excitonic effects that might result in a competition between the two minima in the MoSe$_2$-CB to exhibit the lowest energy. Yet, for the HS, the charge transfer across the heterojunction results in a significant doping level of excess electrons in the MoSe$_2$-CB as well as holes in the WSe$_2$-VB. In turn, the quasi Fermi level in the MoSe$_2$-CB is expected to be high enough, such that electrons populate both minima, i.e. the one at the $\Sigma$-valley and the one at the K- and K'-points. In turn, we can indeed expect two transitions for interlayer excitons. Although the calculations are done for the individual monolayers, the correct trends should be captured. In particular, hybridization effects in van der Waals HS are known to affect predominantly the electronic bands at the $\Gamma$ point and only minor at the $\Sigma$-valley, while the K- and K'-points are almost unaffected [25].

The lattice constants of MoSe$_2$ and WSe$_2$ are $a_{MoSe2}$ = 3.288 A and $a_{WSe2}$ = 3.280 A [56]. This mismatch gives rise to a Moiré pattern, which can impact the electronic band structure [44] [50]. In turn, the lattice mismatch between MoSe$_2$ and WSe$_2$ and/or a small twist angle between the two monolayers shifts the corners of the Brillouin zone away from each other. Consequently, Coulomb coupled electron-hole pairs located exactly at the CB minimum at K$^{MoSe2}$ and the VB maximum (i.e. at K$^{WSe2}$) are situated out of the light cone as well. In other words, the transition at K$^{MoSe2}$ - K$^{WSe2}$ is also slightly indirect in momentum space but less than the one at $\Sigma^{MoSe2}$ - K$^{WSe2}$. In turn, such bound electron-hole pairs or excitons cannot recombine radiatively in a direct optical transition. They can only recombine if their momentum sum equals the displacement vector between the K$^{MoSe2}$ and the K$^{WSe2}$ corners, as recently proposed in [48]. A direct fascinating consequence of this scenario are interlayer excitons with a finite kinematic momentum [48]. In the investigated HS, the displacement vector is expected to be small but finite and only caused by the lattice constant mismatch since the twist angle of $\pm 1°$ is negligible. As a result, increasing moderately the doping by photoexcitation lifts the quasi Fermi level for electrons $E^e_{Fermi}$ and holes $E^h_{Fermi}$, so that the sum of the quasi Fermi momenta ($k^e_{Fermi} + k^h_{Fermi}$) can get larger than the displacement vector between the K$^{MoSe2}$ and K$^{WSe2}$ corners [48]. In this way, we interpret that the more direct transition K$^{MoSe2}$ - K$^{WSe2}$ can be activated as a function of doping, which is consistent with the observed laser power dependence [cf. Figure 2]. Furthermore, the transition can be thermally activated by increasing the temperature. As mentioned, the second transition $\Sigma^{MoSe_2}$ - $K^{WSe_2}$ is significantly more indirect in momentum space. There, the interaction with phonons is required to fulfill the momentum conservation for a radiative recombination. A suitable phonon branch of the MoSe$_2$ lattice would be given e.g. by LA phonons particularly at the M point coupling electronic states the $\Sigma$- and K-points in reciprocal space [41]. We argue that the LA(M) phonon with an energy of around $\hbar\omega_{phonon} \approx 15$ meV has a high density of state, and in particular at the presence of doping, the corresponding electron-phonon interaction is strong [41]. This assignment of the IXs explains the dominance of the IX$^\nabla$ peak at low laser power, while the more direct transition IX$^\blacktriangledown$ dominates the high power regime. Equally, it is consistent with the generally larger FWHM for the IX$^\nabla$ peak compared to the IX$^\blacktriangledown$ peak due to the required electron-phonon scattering to conserve the overall momentum within the indirect IX$^\nabla$ transition. We note that the FWHM for the IX$^\nabla$ peak seems to be largely independent from the laser power at 3 K [cf. Figure 2(e)]. An underlying electron-phonon interaction for the IX$^\nabla$ transition further explains that the spatial distribution of the IX$^\nabla$ peak seems to be somewhat smoothed out even at the lowest temperatures, while the spatial distribution of the energetically higher IX$^\blacktriangledown$ peak shows inhomogeneities (cf. Figure 5). We point out that this low temperature and low power regime also exhibits the additional narrow emission peaks in the spectral range of the IX$^\blacktriangledown$ [arrows in Figure 2(a)]. Intriguingly, the narrow peaks do not occur at the absolute lowest energy transitions of the excitonic system. In turn, we interpret the peaks to be induced by intrinsic or extrinsic modulations of the potential landscape, i.e. either by a Moiré pattern [50] or by trapped molecules and residues at the hetero-interface [21]. For large excitation densities, such potential minima would be smoothed out by an increasing exciton density, which is consistent with our experimental observations.

The dependence of the emission energy on the power and temperature is a complex interplay between dipolar repulsion, exciton diffusion to potential minima [21] and also to a doping-dependent renormalization of the electronic bands that particularly effects the CB minima at the $\Sigma$- and the K'- and K-points with a different strength [39], as already demonstrated in Figure 1(c). In addition, temperature and many-body phenomena affect the single-particle bands such that an altered strength of the Coulomb interaction can play an important role for different charge carrier and exciton densities as well as temperature regimes. In first order, the temperature and power dependence of IX$^\blacktriangledown$ is interpreted as an interplay between the dipolar repulsion of the IX ensemble and an exciton diffusion. This interplay already results in a blue-shift of the emission energy with increasing exciton density, as is consistent with our data and equally suggested in [21]. In agreement with having indirect IX$^\nabla$ and a more direct IX$^\blacktriangledown$, both the time-resolved and time-integrated data show only a weak polarization dependence (cf. Figure 4). In particular, the involved scattering process with phonons and other charge carriers explain that the transitions IX$^\nabla$ and IX$^\blacktriangledown$ show a rather low polarization dependence. Intriguingly, we observe that the PL lifetimes of the IX$^\nabla$ get shorter for higher powers, while the ones for IX$^\blacktriangledown$ stay the same [cf. Figures 4(e) and 4(f)]. In our understanding, this observation is consistent with an incrementally heated phonon bath, which is involved in the IX$^\nabla$ transition. Finally yet importantly, the fact that the lifetimes of IX$^\nabla$ are significantly longer than the ones of IX$^\blacktriangledown$ indicates that the wave function overlap is reduced for the IX$^\nabla$, as expected for an indirect IX$^\nabla$ compared to a more direct IX$^\blacktriangledown$.

An alternative interpretation for the observed splitting of the IX PL is the formation and dominance of charged excitons, such as trions, with excess charge carriers in the $MoSe_2$-CB or $WSe_2$-VB. A comparison of the power- and temperature-dependent emission intensity and energy for single TMD monolayers [57] and a $MoS_2$ bilayer separated by hexagonal Boron Nitride as tunnel barrier [17], reveals significant differences to the observations shown in Figures 2 and 3. Particularly, the power-dependent emission energy is reported to be opposite for trions and neutral excitons for a $MoS_2$ HS with two monolayers of $MoS_2$ separated by hexagonal Boron nitride. There, the high-energy neutral exciton emission drops in intensity with increasing excitation power, whereas the trion emission increases in intensity with increasing excitation power. As shown in Figure 2, both PL peaks $IX^\blacktriangledown$ and $IX^\triangledown$ increase with increasing power density, and the emission energy of both increase concurrently. As a result, the interpretation of the splitting due to the formation of trions is unlikely.

Another possible explanation of the observed doublet splitting would be the emission from SOC-induced spin-split conduction bands at the K- and K'-points. This explanation is very unlikely because we measure PL lifetimes which are almost identical for co- and cross-polarization [Figures 5(e) and 5(f)]. Only in time-integrated measurements, $IX^\blacktriangledown$ shows a polarization of about 20% at the lowest temperatures, while $IX^\triangledown$ seems to be unpolarized within the given experimental error.

We note that in the high power regime, also non-radiative recombination due to Auger processes might occur, and we cannot exclude that the IXs dissociate and undergo an exciton Mott transition [43]. However, we do not recognize signatures in our data pointing towards an unbound electron and hole plasma in the adjacent layers. Instead, the homogenous filling and lateral distribution of the IXs for high excitation intensities makes the system promising to study the manifold phase diagram of interacting composite bosons that are very stable even at rather high temperatures due to the strong Coulomb interaction of about 200 meV reported for IX in such HS [25]. Finally, we would like to point out that the general dependences and signatures of the IX emission are taken from different spots on the HS. The measurement series is conducted from various cool downs with the HS warmed up to room temperature in between establishing the robustness of the observed signatures.

To conclude, we show a detailed photoluminescence study of interlayer excitons in a $MoSe_2$/$WSe_2$ heterostructure as a function of temperature, excitation power, light polarization, and spatial coordinates. We find two emission peaks with a linewidth of about 50 meV and 40 meV, which are separated by some 30-50 meV in energy. The emission energy of both is ~300 meV below the one of the intralayer excitons in $MoSe_2$- and $WSe_2$-monolayers. For both peaks, we find long lifetimes of several tens of nanoseconds up to 100 ns at 3K. The lifetimes are getting shorter for an increasing bath temperature. The energetically higher peak of the two shows several spikes in the photoluminescence spectrum for a low excitation power and low temperatures. Corresponding spatial maps of the photoluminescence intensity show distinct emission centers at the position of the heterostructure. The spatial and spectral inhomogeneities disappear for higher excitation powers and temperatures. We compare our experimental results to DFT single-particle band structure calculations of $MoSe_2$- and $WSe_2$- monolayers which are doped due to optical excitation. We conclude that there are two sorts of interlayer excitons. The low-energy interlayer exciton is indirect both in real and reciprocal space, while the high-energy interlayer exciton is only indirect in real space.

While writing this manuscript, we got aware of pre-published related work [21]. However, the authors discuss data only for larger laser powers and without mentioning any doublet structure of the interlayer excitons.

**ASSOCIATED CONTENT**

**AUTHOR INFORMATION**

Corresponding Authors

*(U.W.) E-Mail: wurstbauer@wsi.tum.de

*(A.W.H.) E-Mail: holleitner@wsi.tum.de

The authors declare no competing financial interest.

**ACKNOWLEDGMENTS**

We gratefully acknowledge financial support by the Deutsche Forschungsgemeinschaft (DFG) via excellence cluster 'Nanosystems Initiative Munich' as well as DFG projects WU 637/4-1 and HO3324/9-1 and the Munich Quantum Center (MQC). A.S. and F.J. acknowledge financial support by the DFG. The band-structure calculations shown in this publication are directly based on ab initio calculations by G. Schönhoff, M. Rösner and T.O.Wehling (University of Bremen).

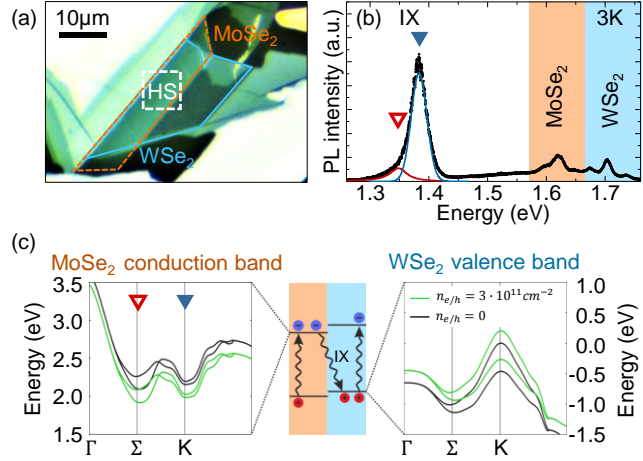

**Figure 1.** Interlayer excitons (IXs) in a MoSe$_2$/WSe$_2$ heterostructure (HS). (a) Optical microscope image of the heterostructure made from monolayers (1L) of MoSe$_2$ and WSe$_2$. (b) Photoluminescence (PL) spectrum taken from the heterostructure area within the dashed square in (a) at a bath temperature of $T_{bath}$ = 3 K for $P$ = 140 µW. At low emission energy, there are two peaks which we define as IX$^\triangledown$ and IX$^\blacktriangledown$. Their energy is lower than the ones for the intralayer excitons in MoSe$_2$ and WSe$_2$. (c) Left panel: computed conduction band of MoSe$_2$ with a charge carrier density of $3 \cdot 10^{11}$ cm$^{-2}$ (green) and without doping. Middle panel: sketch of the type-II band alignment of the MoSe$_2$/WSe$_2$ heterostructure with the IX having the electron and the hole located in MoSe$_2$ and WSe$_2$, respectively. Right panel: computed valence band of WSe$_2$ with a charge carrier density of $3 \cdot 10^{11}$ cm$^{-2}$ (green) and without doping.

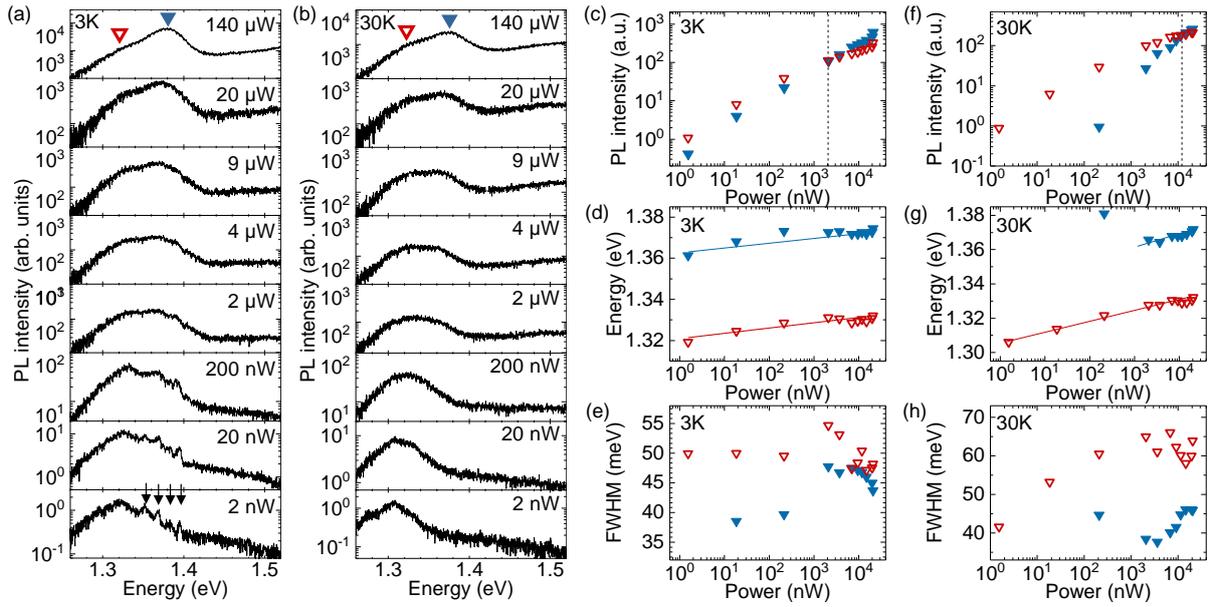

**Figure 2.** Excitation power dependence of the interlayer exciton emission. (a) and (b) Logarithmical waterfall presentation of the PL spectra with the emission peaks IX$^\triangledown$ and IX$^\blacktriangledown$ vs laser power in the range of 2 nW ≤ $P$ ≤ 140 µW for a bath temperature of $T_{bath}$ = 3K (a) and 30K (b). For low $P$ and 3K, we detect emission spikes in the spectral range of the energetically higher IX$^\blacktriangledown$ (arrows), which are not present for the energetically lower IX$^\triangledown$. They equally disappear for high temperatures. (c) to (e) Power-dependence at 3K of the corresponding PL intensity (c), the emission energy (d) and the FWHM (e) with open (filled) triangles representing the data for the IX$^\triangledown$ (IX$^\blacktriangledown$). (f) to (h) equivalent data for 30K. Lines are guides to the eye.

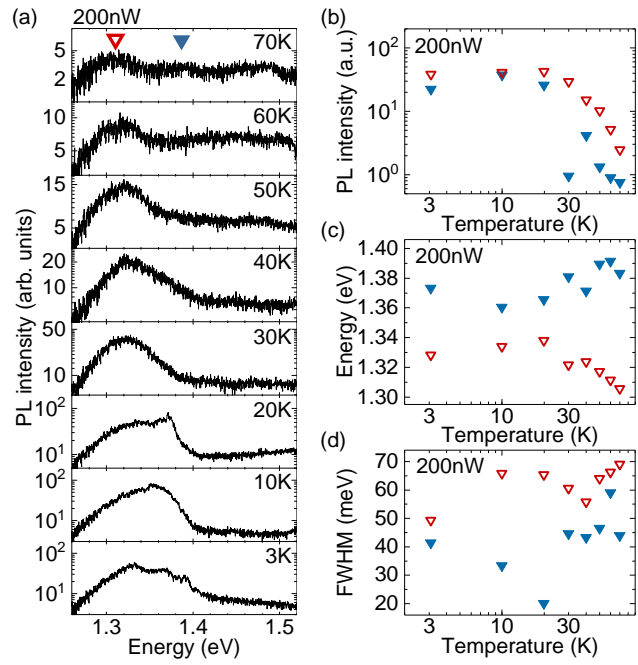

**Figure 3.** Temperature dependence of the interlayer exciton emission. (a) Logarithmical waterfall presentation of PL spectra with the emission peaks IX$^\nabla$ and IX$^\blacktriangledown$ vs temperature in the range of 3K ≤ $T_{\text{bath}}$ ≤ 70 K for a fixed laser power $P$ = 200 nW. (b) to (d) Temperature dependence of the corresponding PL intensity (b), emission energy (c), and FWHM (d) with open (filled) triangles representing the data for the IX$^\nabla$ (IX$^\blacktriangledown$).

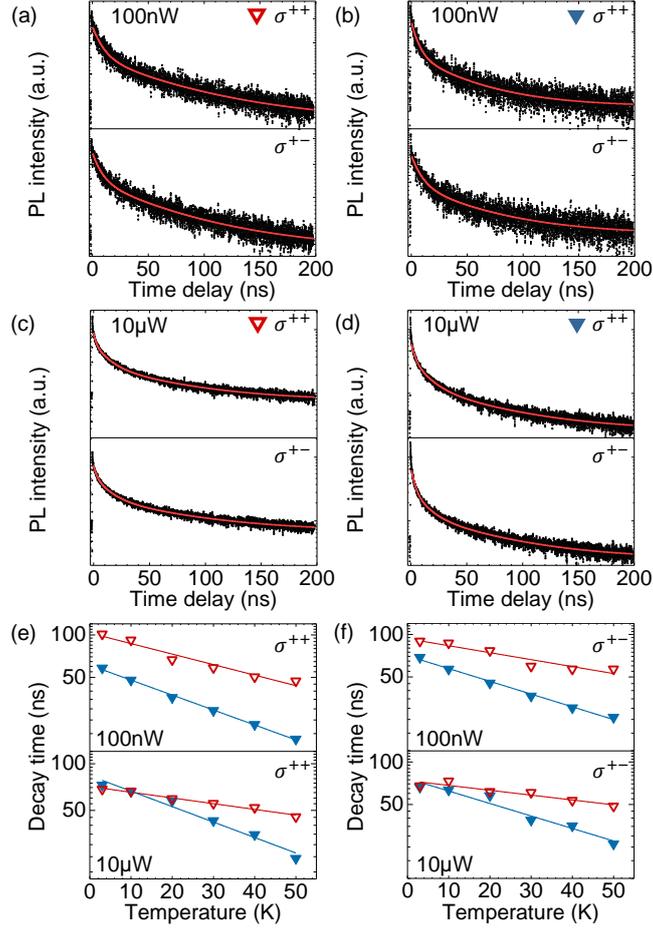

**Figure 4.** Polarized photoluminescence lifetimes of the interlayer excitons. (a) PL signal of the IX$^\nabla$ vs time delay for a co-polarization ($\sigma$++) and cross-polarization ($\sigma$+-) of the excitation and detection, respectively. The data are taken for a low excitation power ($P$ = 200 nW) at 10 K, and they are offset for clarity. (b) Equivalent data for IX$^\blacktriangledown$. (c) and (d): Corresponding data for IX$^\nabla$ (c) and IX$^\blacktriangledown$ (d) for high excitation power ($P$ = 10 µW). Lines are biexponential fits as described in the text. (e) and (f) Temperature dependence of the PL lifetimes at a laser power of $P$ = 200 nW with open (filled) triangles representing the data for the IX$^\nabla$ (IX$^\blacktriangledown$) for co-polarization (e) and cross-polarization (f). Panels (g) and (h) represent equivalent data for $P$ = 10 µW. Lines are guides to the eyes.

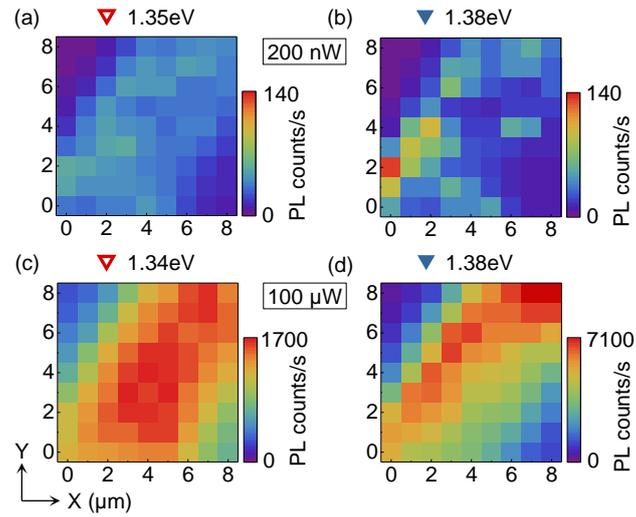

**Figure 5.** Spatial distribution of the IXs' emission signal at low and high excitation power. (a) Spatial PL map of the IX$^\triangledown$ within the area marked with a dashed square in Figure 1(a). The map is taken at a power of $P = 200$ nW. (b) Similar map for the IX$^\blacktriangledown$. (c) and (d): PL map for IX$^\triangledown$ (c) and IX$^\blacktriangledown$ (d) for a high excitation power ($P = 100$ µW). All intensities are given in counts per second (counts/s). The data are measured at 3 K.